\documentclass[11pt]{article}
\usepackage{graphicx}
\usepackage{epsfig}
\usepackage{amsfonts}
\usepackage{amssymb}
\usepackage[dvips]{color}

\newcommand{\insertplot}[5]{\begin{figure}
 \hfill\hbox to 0.05in{\vbox to #5in{\vfill
 \inputplot{#1}{#4}{#5}}\hfill}
 \hfill\vspace{-.1in}
 \caption{#2}\label{#3}
 \end{figure}}
 \newcommand{\inputplot}[3]{
 \special{ps: plotfile #1}
}
\newcounter{fig}

\newcommand{\ee}{\end{equation}}
\newcommand{\eea}{\end{eqnarray}}
\newcommand{\be}{\begin{equation}}
\newcommand{\bea}{\begin{eqnarray}}

\begin{document}

 \title{ Scalarized compact objects in a vector-tensor Horndeski gravity
}

\author{
{\large Y. Brihaye}$^{\dagger}$ \textit{and}
{\large Y. Verbin} $^{\ddagger}$
\\
\\
$^{\dagger}${\small Physique-Math\'ematique, Universit\'e de
Mons, Mons, Belgium}
\\
$^{\ddagger}${\small Department of Natural Sciences, The Open University of Israel, Raanana, Israel}
}
\maketitle
\begin{abstract}
We have discovered a new type of  scalarized charged black holes in a surprisingly simple system: an Einstein-Maxwell-Klein-Gordon field theory where the three fields couple non-minimally through a Horndeski
 vector-tensor term. In addition to hairy charged black holes, this system exhibits Horndeski-Reissner-Nordstrom solutions and ordinary Reissner-Nordstrom ones which bifurcate to the scalarized solutions. There exist also vector-tensor black hole solutions which possess central curvature singularities although their metric components are finite there, and also solutions with naked singularities producing regular metric components everywhere around them. We analyze the solutions and present their main features.
 \end{abstract}
\section{Introduction}\label{introduction}
{In the gigantic effort aimed to understand the dark matter and dark energy
problems of the Universe, numerous extensions of General Relativity (GR) have been
studied in the last few decades. Among these generalized gravities the extensions of the
minimal Einstein-Hilbert lagrangian by scalar and/or vector fields play an important role.
With the main objective of maintaining field equations of the second order in field derivatives,
the works of G. Horndeski \cite{Horndeski:1974wa,Horndeski:1976gi} provoked a
considerable revival of interest in the last years.
Apart from cosmological considerations, the non-minimal geometric gravities of space-time
extended by extra fields lead to the possibility to evade the limitation of
classical solutions due to no-hair theorems \cite{no_hole_old1,no_hole_old2,no_hole_new} valid
in the Einstein-Hilbert theory. A rich set of new compact objects -- hairy black holes (HBHs), boson stars and wormholes --
can be constructed in some extended gravities.
But very surprisingly, black holes with scalar hair were found to exist even with minimal coupling only: Attracting a lot of attention, a family of hairy black holes was first obtained in \cite{Herdeiro:2014goa} within the Einstein gravity minimally coupled to a complex scalar field.
In this case, the  no-hair theorems \cite{no_hole_old1,no_hole_old2,no_hole_new} are bypassed by the
rotation of the black hole and  the synchronization of the spin of the black hole with the (internal) angular frequency of the scalar field.
Reviews on the topic of black holes with scalar hairs
are provided e.g. in \cite{Herdeiro:2015waa,Sotiriou:2015pka,Volkov:2016ehx}.
The scalar-tensor theory of \cite{Horndeski:1974wa}
was revived in the context of Galileon theory \cite{galileon}
and its extensions  in  \cite{Deffayet:2011gz}.
The general lagrangian of scalar tensor gravity is very rich and special cases were studied
by several authors. In many of these studies  the  Gauss-Bonnet  geometric invariant  plays a central role;
although in four dimensions the Gauss-Bonnet term is a pure divergence, it is rendered non trivial
by a coupling to the scalar field through a function $H(\phi)$.
The truncation of the Galileon  theory to lagrangians
admitting a shift symmetric scalar field was worked out by Sotiriou and Zhou
\cite{Sotiriou:2013qea} with a linear function $H(\phi)$.
Static spherically-symmetric hairy black holes were constructed numerically and perturbatively
in \cite{Sotiriou:2014pfa}.
Abandoning the hypothesis of shift symmetry,  several groups \cite{Doneva:2017bvd,Silva:2017uqg,Antoniou:2017acq}
considered during the past couple of years, new types of  non-minimal coupling terms  between a scalar field and specific geometric invariants (essentially the Gauss-Bonnet term).
In these models  the occurrence of hairy black holes
results from an unstable mode associated to the scalar field equation
(a generalized Klein-Gordon equation)
in the background of the underlying metric (the probe limit).
The coupling constant characterizing  the interaction between the scalar field
and the geometric invariant  plays a role of a spectral parameter of the linear equation.
We use to say that the hairy black holes appear through a
{\it spontaneous scalarization}   for a sufficiently large value of the coupling constant.
Away from black holes, it was demonstrated in \cite{Kanti:1995vq} that the energy momentum tensor
associated with a  GB term  also violate the energy condition, opening the door for constructing wormhole
in Einstein-Gauss-Bonnet models coupled to normal matter field.
Such objects were indeed constructed in \cite{Kanti:2011yv} and in \cite{Antoniou:2019awm} where the properties,
domain of existence and stability were studied in detail for some choices of the coupling function $H(\phi)$.}
In a series of recent papers \cite{Herdeiro:2018wub,Fernandes:2019rez,Herdeiro:2019yjy},
other types of extensions of the Einstein-Hilbert-Maxwell lagrangians
have been presented  and families of charged hairy black holes have been constructed
with a non-minimal coupling of the scalar field to the electromagnetic Maxwell field.

So far, up to our knowledge, no analogous investigation was done for the vector-tensor Horndeski theory \cite{Horndeski:1976gi}
extended by a scalar field.
The original vector-tensor theory proposed in \cite{Horndeski:1976gi} is characterized by a single interacting term between the geometry and the vector field with a single coupling constant.
The generic static, spherically symmetric solutions are essentially
deformations of the Reissner-Nordstrom solutions. The electric version of these Horndeski-Reissner-Nordstrom (HRN) black holes were studied in \cite{Horndeski-HRN1978} and in much more detail in \cite{muller}.

In this paper we reconsider the Horndeski vector-tensor lagrangian and we extend it by a real scalar field coupled, again, to the Horndeski interaction term by means of a coupling function $H(\phi)$. In this new type of scalar-vector-tensor theory, we have constructed soliton-like solutions and hairy black holes that bifurcate from the solutions of \cite{muller} at specific values of the coupling constants. Unlike the ordinary Einstein-Maxwell theory where the magnetic solutions are essentially the same as the electric ones, this is not so if the Horndeski term is added and the situation is quite different. Thus we concentrate at present in the electric case and defer the magnetic version to a future publication.

The paper is organized as follows:  in Sec. \ref{TheModel} we present the model and in Sec. \ref{sphsymmsol} the field equations and appropriate boundary conditions. In Sec. \ref{HRNsol} we present the electrically charged black holes of the vector-tensor system and in Sec. \ref{HBH-sec} we arrive at the main subject of this paper, the charged black holes with scalar hair. A summary of the results and some perspectives are the objective of Sec. \ref{summary}.
\\
\\
\section{The model}\label{TheModel}
\setcounter{equation}{0}
We are interested in classical solutions associated with
Einstein-Maxwell-Klein-Gordon lagrangian extended by a non-minimal
coupling term inspired by the work of G. Horndeski.
The action considered is of the form
\be
   S = \int d^4 x \sqrt{-g} \bigg[ \frac{1}{2\kappa} R - \frac{1}{2} \nabla_{\mu} \phi \nabla^{\mu} \phi
	- \frac{1}{4} F_{\mu \nu}F^{\mu \nu} + H(\phi) {\cal I}(g,A)  \bigg]
\label{lagrangian}
\ee
where $R$ is the Ricci scalar, $F_{\mu \nu}$ is the electromagnetic field strength
 and $\phi$ represents a real scalar field.  We use $\kappa = 8 \pi G $ which we later take to be 1 by rescaling. The last term ${\cal I}(g,A)$ is the non-minimal coupling of the vector field to the geometry introduced by Horndeski \cite{Horndeski:1976gi} as a possible interaction term which keeps the field equations of second order.
\be
\label{HorndeskiTerm}
{\cal I}(g,A) = -\frac{1}{4} (F_{\mu \nu} F^{\kappa \lambda} R^{\mu \nu}_{\phantom{\mu \nu} \kappa \lambda}
                          - 4 F_{\mu \kappa} F^{\nu \kappa} R^{\mu}_{{\phantom \mu} \nu}
													+ F_{\mu \nu} F^{\mu \nu} R ) \ .
\ee
Following the spirit of many recent works, the  vector-tensor gravity lagrangian has been augmented
by a real scalar field $\phi$ and
this  scalar is coupled  to the Horndeski interaction term via the coupling function $H(\phi)$. The case of a constant $H(\phi)$ then corresponds to the original vector-tensor Horndeski lagrangian.

Recently, several hairy black holes have been constructed  in other versions of the model of the type (\ref{lagrangian}) above,
mainly by choosing for ${\cal I}(g)$ the Gauss-Bonnet invariant or the Faraday-Maxwell term  ${\cal I}(A) = F_{\mu \nu} F^{\mu \nu}$.
In these works  several forms of the function $H(\phi)$ were  used in order to construct uncharged hairy black holes, see e.g. \cite{Doneva:2017bvd,Silva:2017uqg,Antoniou:2017acq,Antoniou:2017hxj}
as well as charged ones \cite{Brihaye:2019kvj}. Here we will take the ``tripartite'' coupling term and show that it produces a rich and interesting family of compact objects.

\section{Spherically-symmetric solutions}\label{sphsymmsol}
\setcounter{equation}{0}
\subsection{Ansatz}\label{ansatz}
In order to obtain static spherically symmetric solutions we will adopt  a  metric of the form
\be
     ds^2 = - f(r) a^2(r) dt^2 + \frac{1}{f(r)} dr^2 + r^2 d \Omega_2^2
\ee
completed by a spherically-symmetric scalar field $\phi(x^{\mu}) = \phi(r)$
and a spherically-symmetric electric field derived from the vector potential $A_0=V(r), A_{1,2,3} = 0$.

Substituting the ansatz in the field equations, the  system can be reduced to a set of four  non linear differential
equations for the functions $f(r), a(r), V(r)$ and $\phi(r)$ (with the notation $H'(\phi )=d H /d\phi $):

\bea
 \left( r^2 a f \phi ' \right)'+\frac{ H'(\phi ) (1-f) \left(V'\right)^2}{a}&=&0 \label{FEqScalar} \\
 \left[\frac{\left(r^2 +2 H(\phi )(1-f)\right)V' }{a}\right]'&=&0   \label{FEqA0} \\
1-f - r f' &=&\frac{\kappa  r^2}{2}  \left[f \phi '^2 +\frac{ \left(r^2+ 2  H(\phi ) (1-f)\right)(V')^2}{r^2 a^2}\right] \label{FEqf} \\
\frac{r a'}{a}&=&\frac{\kappa  r^2}{2} \left(\phi '^2 + \frac{2 H(\phi ) (V')^2}{r^2 a^2}\right) \label{FEqa}
\eea

The potential  $V(r)$ can be eliminated by using the Maxwell equation (the full equations depend on $V'$ only), leading to
\be
V'(r) = \frac{ Q a(r)}{ r^2 + 2 H(\phi(r) )(1-f(r))}
\label{ElField}
\ee
where $Q$ is the integration constant which we interpret as the electric charge of the solution. Without loss of generality, we will assume $Q \geq 0$. With this elimination,
the three final equations are of the first order for the functions $f(r)$, $a(r)$ and of the second order for $\phi(r)$. They depend on the charge parameter $Q$ and the shape of the scalar coupling function
$ H(\phi)$. For the initial study of this work we take the simple choice
\be
\label{coupling}
               H(\phi) = \gamma + \alpha \kappa \phi^2
\ee
where $\alpha$, $\gamma$ are independent coupling constants. This parametrization contains the case of $\alpha = 0$ which corresponds to the  original vector-tensor Horndeski lagrangian and $\gamma = 0$ with $\alpha > 0$ which as we will see, produces scalarized black holes.  The system depends therefore on the coupling constants $\alpha$ and $\gamma$ and the charge parameter $Q$. These charged scalarized black holes constitute a new family of such solutions  supported by  a non-minimal coupling to electromagnetism  through the term (\ref{HorndeskiTerm}).   Charged scalarized black holes are already known to exist \cite{Herdeiro:2018wub} in the Einstein-Maxwell-Klein-Gordon system  with a different  non-minimal  scalar-electromagnetic coupling and also in another generalization of the theory with an additional $(F_{\mu\nu}{^*}F^{\mu\nu})^2$ term \cite{Herdeiro:2019yjy}.

\subsection{Boundary conditions and asymptotics}\label{BC+Asymp}
For a fixed choice of $\alpha, \gamma$ and $Q$,
four  conditions on the boundary need to be fixed to specify a solution.
Because we  look for localized, asymptotically flat solutions, we require
\be
\label{cond_infty}
   a(r \to \infty) = 1 \ \ \ \ , \ \ \ \  \phi(r \to \infty) = 0 \ \ \ \ , \ \ \ \ m(r \to \infty) = M
\ee
where we define the mass function by
\be
\label{f and M}
   f(r) = 1 - \frac{2 m(r)}{r}
 \ee
and  $M$ is  a positive constant. For black hole solutions, imposing a regular horizon at $r=r_h$ needs $f(r_h)=0$. The equation of the scalar field is then singular in the limit $r\to r_h$; obtaining regularity requires a very specific relation between the values $\phi(r_h)$ and $\phi'(r_h)$. Examining the equations the conditions for a regular scalarized black hole at $r_h$ turn out to be~:
\be
\label{cond_regular}
   f(r_h) = 0 \ \ , \ \   \phi'(r_h) = \frac{2 Q^2  H' }{r(2 H + r^2)(\kappa Q^2 - 2 r^2 - 4 H )}|_{r=r_h} \ \ .
  \ee

The boundary value problem is then fully specified by (\ref{cond_infty}) and (\ref{cond_regular}).
Existence of an hairy black hole will imply in addition
setting a value $\phi(r_h) \neq 0$ for the scalar field;
this clearly leads to a relation between the constants $\alpha$, $\gamma$, $Q$ and $\phi(r_h)$. Stated differently, for any $\alpha$ and $\gamma$, the value of $\phi(r_h)$ fixes the values of the charge $Q$ or vice-versa. In addition to the electric charge $Q$,  the solutions will be further characterized by the mass $M$, scalar charge $D$, the temperature $T_H = a(r_h) f'(r_h)/(4 \pi)$ and the horizon area $A_H = 4 \pi r_h^2 $. Using the field equations for $r=r_h$ we find the explicit equation for the temperature:
\be
\label{Temperature}
  T_H =\frac{a(r_h) (2 r_h^2 + 4 H(\phi_h)-\kappa Q^2)}{8\pi r_h (r^2_h + 2 H(\phi_h))}   \ \ .
  \ee

\section{HRN black holes and naked singularities
}\label{HRNsol}
\setcounter{equation}{0}

\subsection{Reissner-Nordstrom solutions}\label{RNBHs}
In the absence of the scalar field and the non-linear Horndeski term (i.e. $\phi(r)=0$ and $\gamma=0$), the generic solutions are the Reissner-Nordstrom (RN) black holes
\be\label{RNSolutions}
      f(r) = 1 - \frac{r_h^2+ \kappa Q^2/2}{r_h r} + \frac{\kappa Q^2}{2 r^2} \ \ , \ \ a(r) = 1 \ \ , \ \ V(r) =  - \frac{Q}{r}
\ee
The outer horizon is situated at $r=r_h$ and is simple for $0 \leq \kappa Q^2 < 2 r_h^2$.
The mass and temperature of these black holes are respectively $M= (2r_h^2 + \kappa Q^2)/(4 r_h)$ and $T_H = (2r_h^2 - \kappa Q^2)/(8\pi r_h^3)$.
The horizon   becomes double for $\kappa Q^2 = 2 r_h^2$  corresponding to
an extremal  black hole. For the future discussion, let us point out that
these extremal black holes have $M = r_h = \sqrt {\kappa / 2}\;Q$ so that $\sqrt {\kappa} Q/M= \sqrt 2$, irrespectively of $r_h$.

\subsection{HRN black holes}\label{HRNBHs}
\begin{figure}[b!!]
\begin{center}
{\includegraphics[width=8cm, angle = -00]{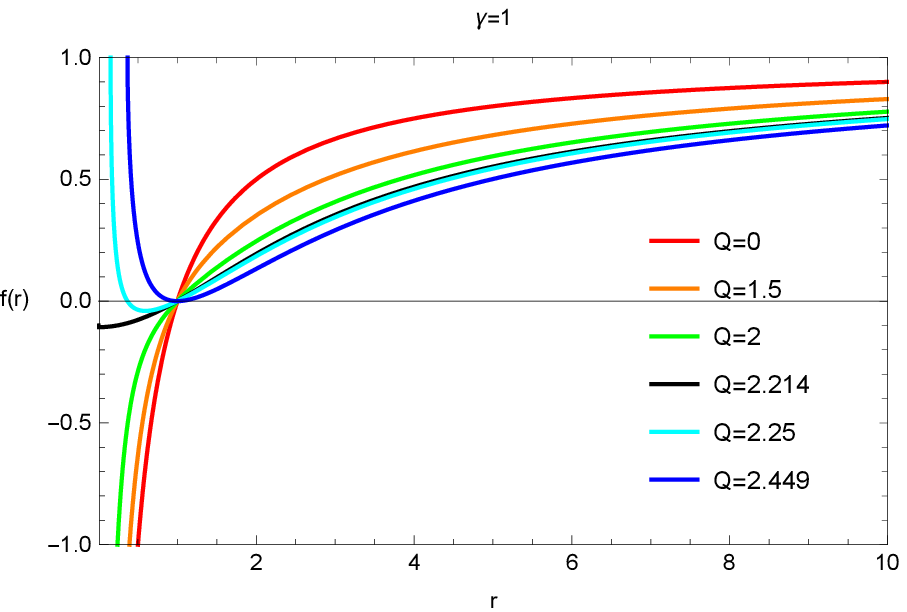}}
{\includegraphics[width=8cm, angle = -00]{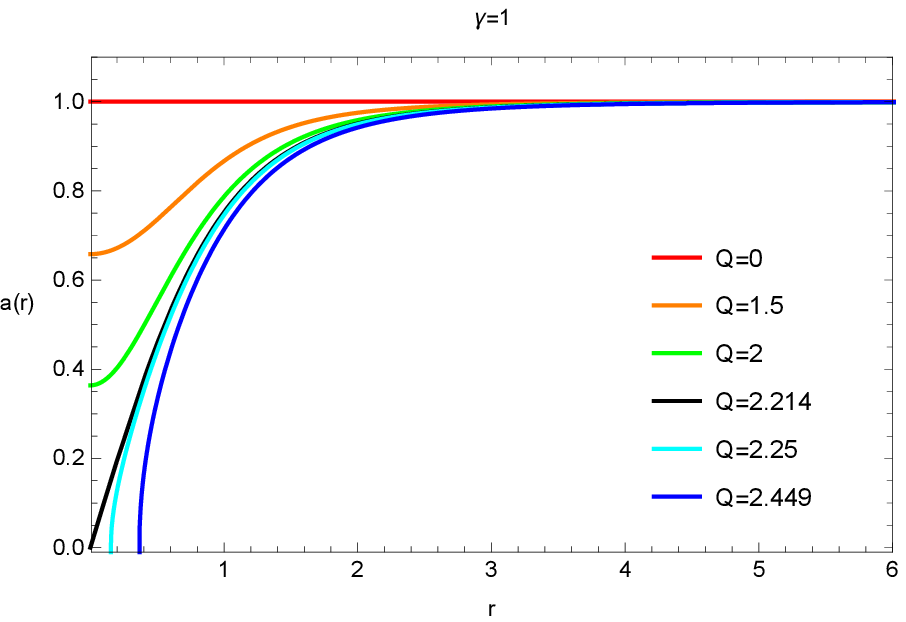}}
\end{center}
\caption{Several profiles of HRN black holes with $Q$ from $Q=0$ (Schwarzschild) to the maximal possible $Q=\sqrt{6}$ for $\gamma=1$ (see text). The value $Q=2.214$ is a special solution with finite $f(0)$.  Note that the figure includes the region inside the event horizon which is for all solutions at $r=1$.
\label{HorndeskiBH-Profiles}
}
\end{figure}
If the non-minimal coupling of the Horndeski term is present, but the scalar field is still absent (i.e. $\phi(r)=0$ and $ H(\phi) = \gamma$ - see Eq. (\ref{coupling})), the solutions are deformations of the Reissner-Nordstrom solutions \cite{muller}. To first order in the non minimal coupling constant $\gamma$, we found the following asymptotic behaviour
\be\label{Asymptotics1-HRN}
   f(r) = 1 - \frac{2 m(r)}{r} \ \ , \ \ m(r) = M - \frac{\kappa Q^2}{4 r} + \frac{\gamma \kappa Q^2 M}{4 r^4} + \dots
\ee
\be\label{Asymptotics2-HRN}
       a(r) = 1 - \frac{ \gamma \kappa Q^2}{4 r^4} + \dots \ \ , \ \ V(r) =  - \frac{Q}{r} + \frac{\gamma M Q}{r^4}+ \dots
\ee
These solutions were already studied in ref. \cite{muller}, so we will be brief here, although our method is somewhat different and we extend our analysis beyond ref. \cite{muller}. We solved the system by using the numerical routine COLSYS \cite{COLSYS} as will be described in more details in Sec. \ref{physicalparameters}. In the numerical construction of solutions
 we took advantage of the freedom to rescale  the radial coordinate $r$ as  $r/r_h \rightarrow r$ so the event horizon is located at $r=1$ for all black hole solutions. The electromagnetic potential $A_{\mu}$ the charge $Q$ and the mass $M$ are rescaled accordingly to become dimensionless in such a way that we may set $\kappa= 8 \pi G = 1$. The coupling constant $\gamma$ becomes dimensionless as well. In what follows we will not distinguish in most cases between rescaled and non-rescaled quantities. It can be inferred from the context which of those we mean.

We present in Fig. \ref{HorndeskiBH-Profiles} several generic profiles of these Horndeski-Reissner-Nordstrom black holes. Three types of solutions are easily recognized: Schwarzschild-like which exist in the region of small charge starting from the $Q=0$ solution and RN-like which exist for larger values of charge. In between there exists a special solution with a finite value of $f(0)$ dividing between these two families.  The black hole solutions for a given value of $\gamma$ are characterized by the electric charge as shown clearly in Fig. \ref{HorndeskiBH-Profiles}. For any $\gamma$ the charge is limited to a finite interval $0\leq Q \leq Q_{max}(\gamma)$ which exceeds the RN limit of $\sqrt {\kappa} Q/M= \sqrt 2$ mentioned above.  Fig. \ref{HorndeskiBH-Profiles} shows the situation for $\gamma = 1$ for which  $Q_{max}(1) = \sqrt{6}$ as we will explain in what follows. See Eq. (\ref{Q-Interval-HRNBH}) below. The other special value $Q=2.214$ which corresponds to the special solution with finite $f(0)$ marks the border between the two families. We will see in the next subsection that there is a whole family of solutions with finite $f(0)$ -- all characterized by $a(0)=0$.

\begin{figure}[b!]
\begin{center}
{\includegraphics[width=7.4cm, angle = -00]{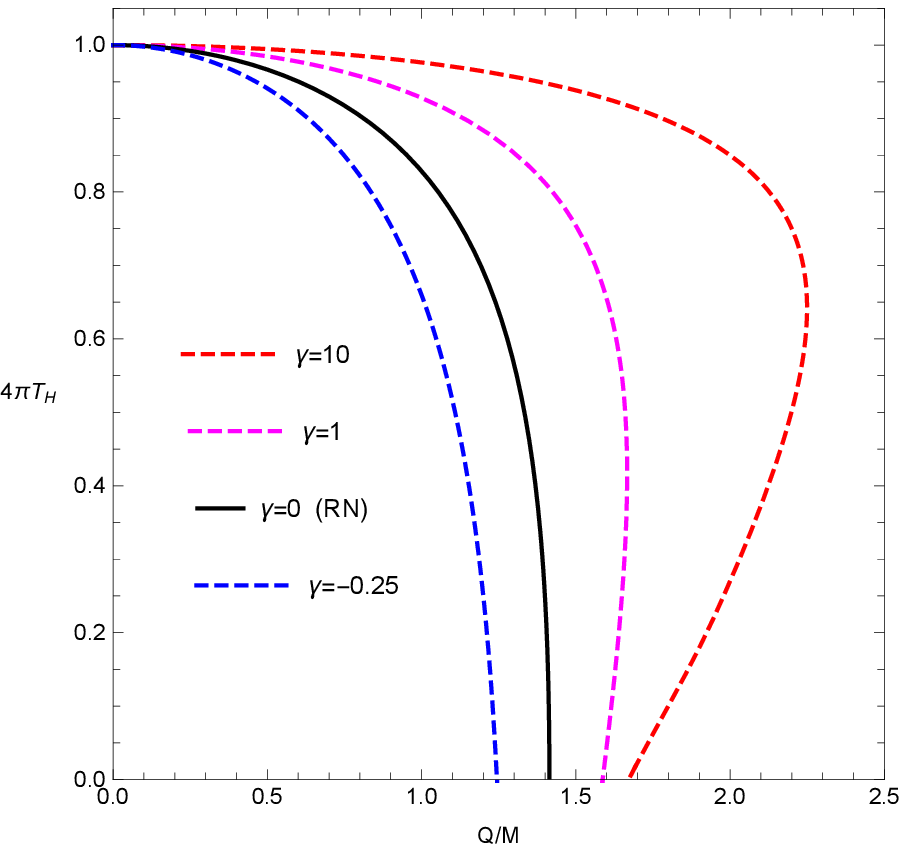}}
{\includegraphics[width=8cm, angle = -00]{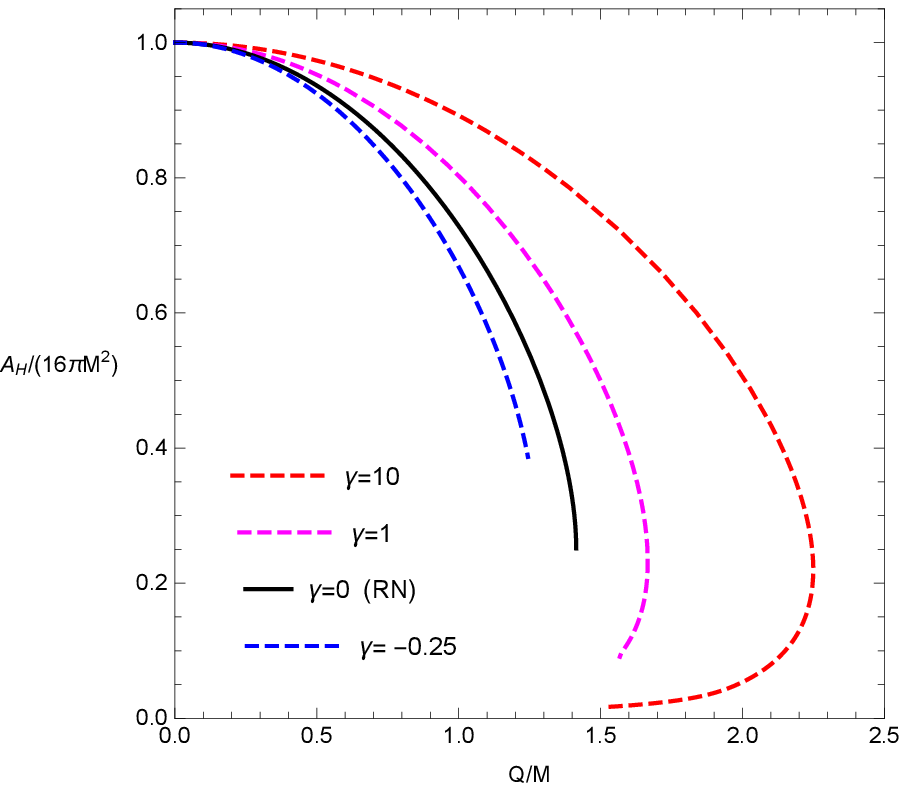}}
\end{center}
\caption{The temperature and horizon area as function of  $Q/M$  for the RN (solid lines) and HRN black holes with $\gamma=-0.25, 1, 10$ (dashed lines). Left: Temperature.  Right: The ratio  $A_H/ 16 \pi M^2$.
\label{ThermodynData-HRNbh}
}
\end{figure}

The domain of existence of these solutions in the $(\gamma, Q)$ parameter-space
can be determined from the temperature formula Eq. (\ref{Temperature}).
In the absence of a scalar field it simplifies to
\be
T_H = \frac{a(r_h)}{4 \pi r_h} \left(1 - \frac{\kappa}{2} \frac{Q^2}{r_h^2 + 2 \gamma}\right) \ \ ,
 \label{TH-HRNBH}
\ee
leading to the condition
\be
         Q^2  \leq \frac{2}{\kappa}(r_h^2 + 2 \gamma)  \ \ {\rm if} \ \ \  r_h^2 + 2 \gamma > 0
 \label{Q-Interval-HRNBH}
 \ee
and no condition on $Q^2$ if $r_h^2 + 2 \gamma < 0$ (this case was not considered, however, because
the ``electric  field'' (\ref{ElField}) then manifestly presents  a singularity).
For a fixed $\gamma$ such that  $r_h^2 + 2 \gamma > 0$, a maximal value of $Q$,
(for instance $Q^2_{max} = 2(1+2 \gamma)$ with our conventions) clearly exists and corresponds to an extremal solution with $T_H = 0$. In Figure \ref{ThermodynData-HRNbh} we demonstrate the behavior of the temperature $T_H$ and horizon area $A_H$ as a function of the charge to mass ratio $Q/M$. This figure summarizes an analysis that goes beyond ref. \cite{muller} and reveals in particular the occurrence of two solutions corresponding  to the same $Q/M$-ratio. It also shows that for $\gamma > 0$ the RN limit
$Q/M = \sqrt{2}$ can be exceeded; the non minimal coupling therefore allows for overcharged (with respect to RN) black holes.
For these solutions, the mass increases monotonically with $Q$. Finally we notice that $0 < a(r_h) \leq 1$ for $\gamma > 0$ and $a(r_h) \geq 1$ for $ r_h^2/2 < \gamma <  0$.

\subsection{HRN  solutions of a second kind: $a(0)=0$}\label{HRNRegular}
 In addition, we have found a new family of vacuum solutions of this system where the metric components are finite at $r=0$ unlike the ordinary HRN solutions.
 These solutions solve as before  Eqs (\ref{FEqA0})-(\ref{FEqa})) with $\phi(r) = 0$ and $H(\phi)=\gamma$, but are characterized by $a(0)=0$ unlike the HRN ones which have $a(0)>0$. Therefore, these ``regular'' solutions satisfy the following conditions at the origin:
 \be
 \label{BC-HorndeskiRegular}
 m(0) = 0 \ \ \ \ , \ \ \ \  f(0) = 1 - 2m'(0) =  1- \sqrt{\frac{\kappa}{\gamma}}\frac{Q}{2}\ \ \ \  , \ \ \ \   a(0) = 0 \ \ \ \
\ee
 Although  the metric components of these solutions are finite at the origin, there is actually a curvature singularity at $r=0$. The singularity is evident as the Ricci and Kretschmann invariants diverge as $r\rightarrow 0$.
 \begin{figure}[b!!]
\begin{center}
{\includegraphics[width=8cm, angle = -00]{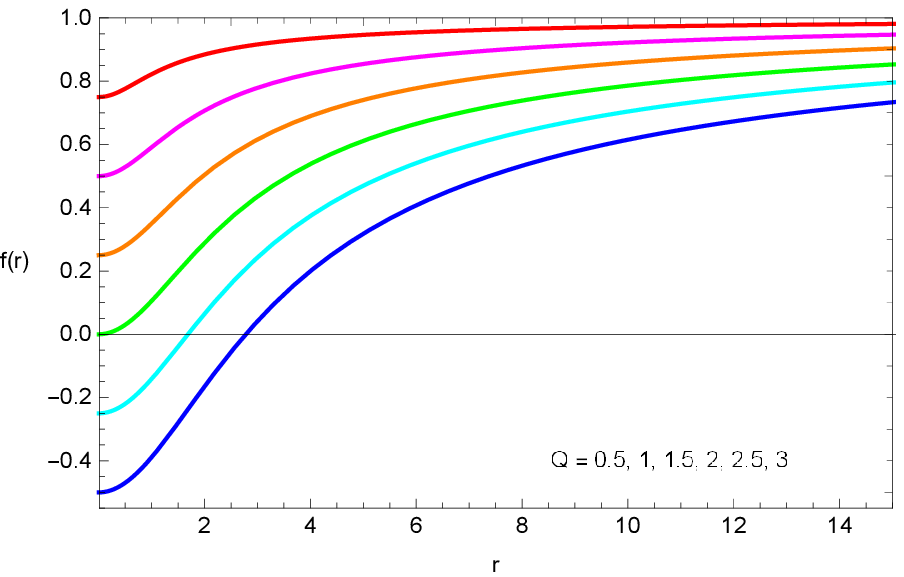}}
{\includegraphics[width=8cm, angle = -00]{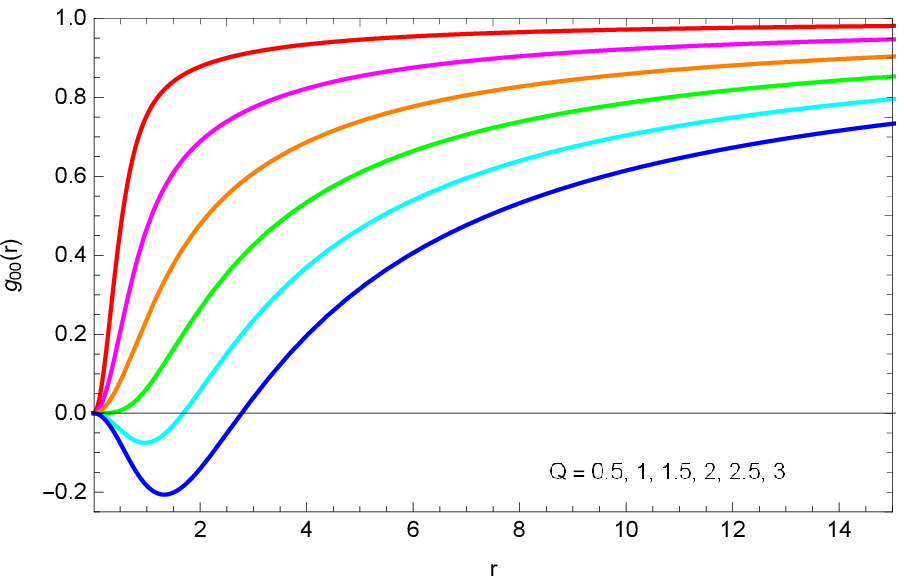}}
\end{center}
\caption{Several profiles of $a(0)=0$ solutions of the Horndeski vector-tensor system for $0<\bar{Q}<3$. The curves with the lower $f(0)$ correspond to the larger $\bar{Q}$ - values.
\label{HorndeskiRegular-Profiles}
}
\end{figure}
These $a(0) = 0$ solutions are also parametrized by the rescaled electric charge  $\bar{Q} = \sqrt{\kappa / \gamma} \,\, Q$  which is the only free parameter for this kind of solutions. They are divided into two subclasses: the first contains the solutions for which $f(r)>0$ everywhere, that is solutions which exhibit a naked singularity. These solutions exist in the charge interval   $0<\sqrt{\kappa / \gamma} \,\, Q<2$ which by Eq (\ref{BC-HorndeskiRegular}) corresponds to $f(0)>0$. The solutions of the second subclass are actually BH solutions which hide their singularity by a horizon where $f(r_h)=0$ and turns negative but finite for $r<r_h$. These solutions exist for $\sqrt{\kappa / \gamma} \,\, Q>2$ and they were actually mentioned briefly before as the special solutions which divide between the Schwarzschild-like and RN-like HRN black holes. Fig. \ref{HorndeskiRegular-Profiles} depicts the metric functions $f(r)$ and $g_{00}(r)=f(r)a(r)^2$ of the solutions of both subclasses.

\section{Black holes with scalar hair}\label{HBH-sec}
\setcounter{equation}{0}

\subsection{Physical parameters}\label{physicalparameters}
The main result of this work is the construction of the scalarized HRN black holes.  We take the coupling function of Eq. (\ref{coupling}) with $\gamma=0$, i.e. $H(\phi)=\alpha \kappa \phi^2$.

These hairy black holes   can
 be characterized by their mass $M$, electric charge $Q$ and scalar charge $D$, related respectively to the asymptotic decay
of the functions $m(r)$ (or$f(r)$), $V(r)$ and $\phi(r)$. The asymptotic behavior is:
\bea\label{Asymptotics1-HRN-Scalar} \nonumber
      m(r) &=& M - \frac{\kappa (Q^2+D^2)}{4 r} - \frac{\kappa M D^2}{4 r^2} + \frac{\kappa D^2(\kappa Q^2 - 8M^2)}{24 r^3}-  \frac{ \kappa M D^2( 48 M^2-12 \kappa Q^2- \kappa D^2 )}{96 r^4} + \dots \\
      a(r) &=& 1 - \frac{ \kappa D^2}{4 r^2}   - \frac{2\kappa M D^2}{3 r^3}   -  \frac{ \kappa D^2( 48 M^2-4 \kappa Q^2- 3\kappa D^2 )}{32 r^4} + \dots
\eea
\bea\label{Asymptotics2-HRN-Scalar}
     V(r)&=&  - \frac{Q}{r} + \frac{\kappa Q D^2}{12r^3} + \frac{\kappa M Q D^2}{6r^4}+ \dots  \\ \nonumber
     \phi(r) &=& \frac{D}{r} +\frac{M D}{r^2} + \frac{D( 16 M^2-2 \kappa Q^2- \kappa D^2 )}{12 r^3}   +  \frac{M D ( 12 M^2-3 \kappa Q^2- 2\kappa D^2 )}{6 r^4} \dots  \ \ .
\eea
The  temperature $T_H = a(r_h) f'(r_h)/(4 \pi)$ further characterizes the solutions. Note that the scalar charge $D$ is an independent charge and is not fixed by $M$ and $Q$.

 The non-minimal coupling constant $\alpha$ appears explicitly only in the higher order terms of the asymptotic expansion, not included in Eqs. (\ref{Asymptotics1-HRN-Scalar})--(\ref{Asymptotics2-HRN-Scalar}). It appears at order $1/r^6$ in the expansion for $a(r)$ and in even higher terms in the others. The coefficients are quite lengthy and we didn't write them down. The effect of this coupling constant
 is of course evident from non-vanishing scalar charge. This is also the reason why the limit $D\rightarrow 0$ does not lead from Eqs. (\ref{Asymptotics1-HRN-Scalar})--(\ref{Asymptotics2-HRN-Scalar}) to Eqs. (\ref{Asymptotics1-HRN})--(\ref{Asymptotics2-HRN}).

Since the non-linear equations do not admit closed form solutions, we solved the system by using the numerical routine COLSYS \cite{COLSYS}.
It is based on a collocation method for boundary-value differential equations and on damped Newton-Raphson iterations. The equations are solved with a mesh of a few hundred points and  relative errors of the order of $10^{-6}$. Various checks on the solutions were performed by obtaining them independently using the MATHEMATICA package.

In the numerical construction of solutions
 we took advantage of a rescaling of the radial coordinate $r$ and of the matter fields $\phi, A_{\mu}$
to set $\kappa= 8 \pi G = 1$ and $r_h=1$. The coupling constant $\alpha$ becomes dimensionless as well. Accordingly, the regularity conditions (\ref{cond_regular}) take the form:
\be
\label{RegCondDimensionless}
   f(1) = 0 \ \ , \ \   \phi'(1) = \frac{4 Q^2 \alpha \phi(1) }{(2 \alpha \phi(1)^2 + 1)( Q^2 - 2 - 4 \alpha \phi(1)^2 )}
\ee
where we used the further simplification to construct black hole solutions in the case $\gamma=0$ only. We will keep the assumption $\gamma=0$ for the rest of this paper.


\begin{figure}[t!]
\begin{center}
{\includegraphics[width=8cm, angle = -00]{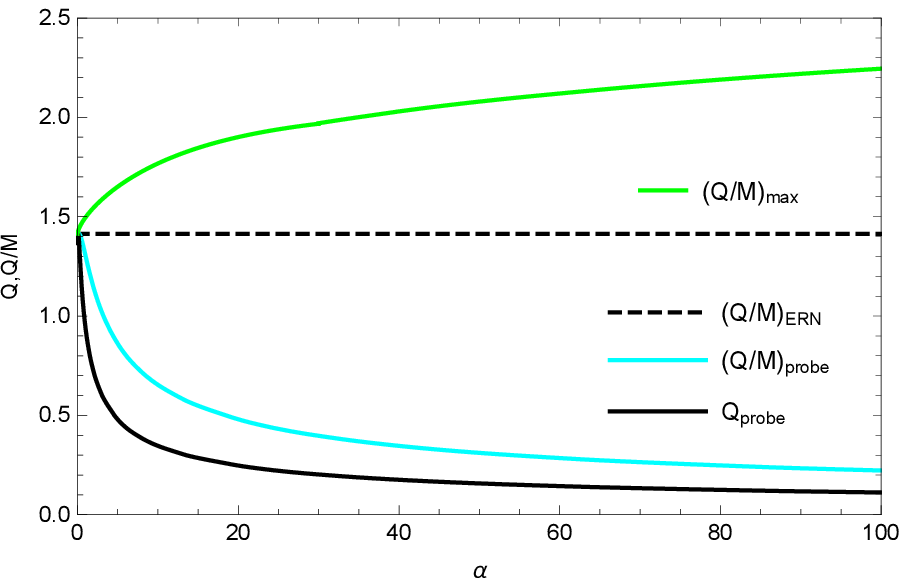}}
{\includegraphics[width=8cm, angle = -00]{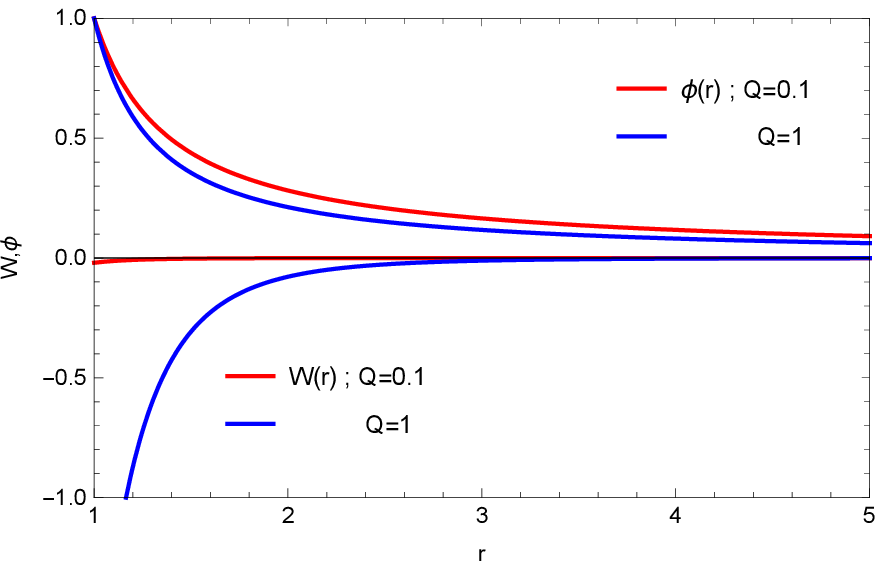}}
\end{center}
\caption{Left: The domain of existence of spherical hairy  black holes in the ($\alpha$ , $Q/M$) plane: they exist between the blue and the red $Q/M$ lines. The line $(Q/M)_{ERN}$ = $\sqrt{2}$ for the extremal RN solution is added as a reference. The curve of $Q$ vs. $\alpha$ of the probe limit corresponds to the line along which exist regular, localized, nodeless $\phi(r)$.
Right: The ``potential'' $W(r)$ and the eigenmode $\phi(r)$ (normalized to $\phi(r_h)=1$) for two values of the charge $Q$.
\label{alpha_q_crit}
}
\end{figure}
\subsection{Probe limit}\label{probelimit}
Before attacking the construction of hairy  black holes, it is useful to
study the equation of the scalar field in the background of a RN black hole, that is to say
that we first solve the Klein-Gordon equation sourced by the non-minimal coupling term.
Since we assume  $H(\phi)=\alpha \kappa \phi^2$ the equation is linear and of the form~:
\be
\label{probe}
     (r^2 a f \phi')' =  \alpha W \phi \ \ , \ \ W(r) \equiv - 2 (1-f(r)) (V')^2/a(r).
\ee
It turns out that for values of $Q$ such that $0 \leq Q \leq \sqrt 2$ this equation admits one critical value of
the coupling constant, say $\alpha(Q)$, for which the equation admits a regular, localized, nodeless solution for $\phi$.
(Note:  higher critical values of $\alpha$
exist with eigenfunction $\phi$ presenting one or several modes exist as well
but we do not report them here).
The relation between  $\alpha$ and $Q$ is presented on the left  side of Fig. \ref{alpha_q_crit};
the profile of the potential $W(r)$ and the corresponding eigenfunction $\phi(r)$ are shown on the right side of the figure
for two values of $Q$ and setting $r_h = 1$.
As known from previous studies, the existence of such solutions constitutes
an evidence for a tachyonic instability of the RN solution in the presence of a scalar field and leads to a bifurcation of hairy  black holes
from the RN solutions. We use to say that the hairy solutions ``appear spontaneously''  $Q > Q(\alpha)$.

\subsection{Hairy black holes}\label{HBH-subsec}

\begin{figure}[b!]
\begin{center}
{\includegraphics[width=8cm, angle = -00]{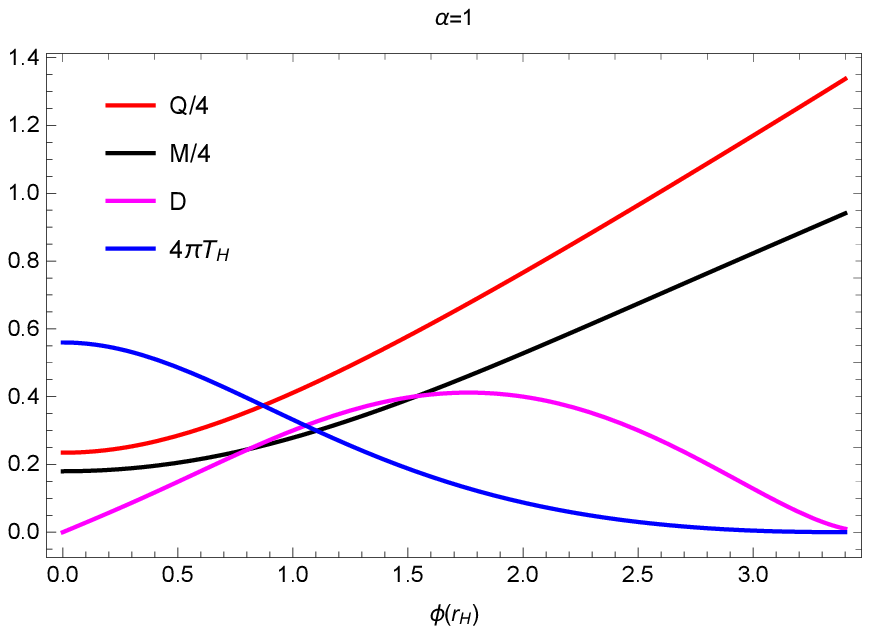}}
{\includegraphics[width=8cm, angle = -00]{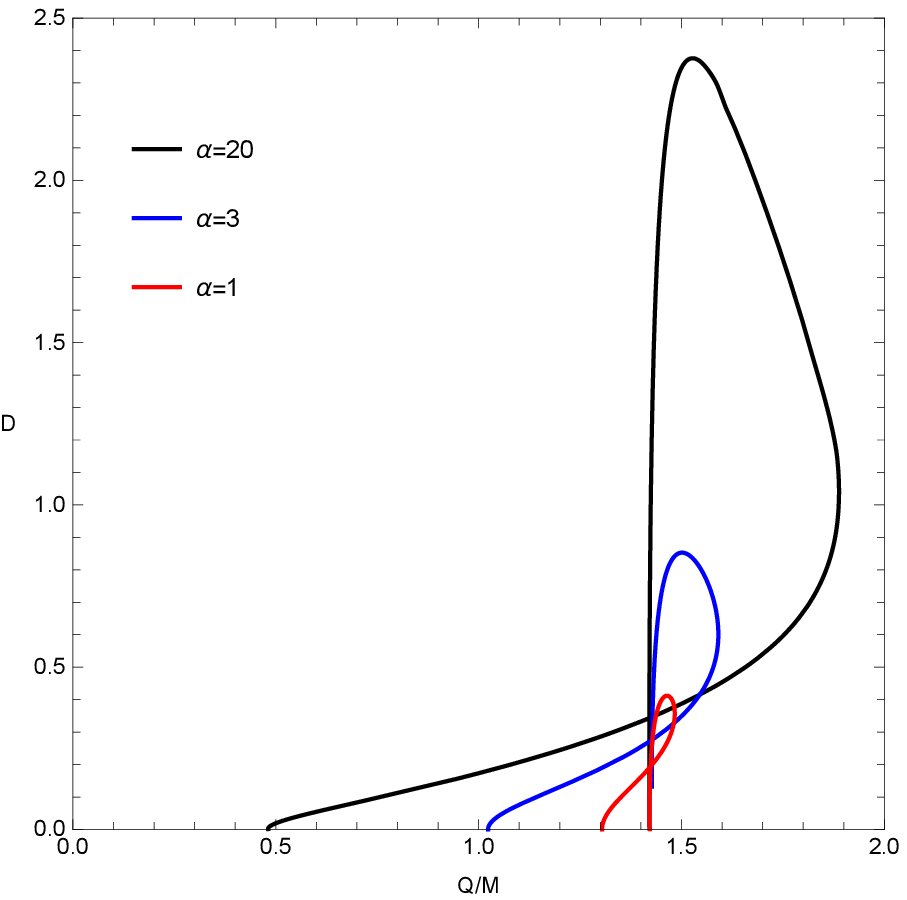}}
\end{center}
\caption{Left: The dependence on the horizon value of the scalar field of  mass, electric charge, scalar charge and temperature of the HBH corresponding to $\alpha = 1$.
Right: Three curves of HBH scalar charge $D$ vs. $Q/M$ for three values of $\alpha$: $\alpha=1, 3, 20$.
\label{data_ga_3_bis}
}
\end{figure}

Since we found regular, localized, nodeless solutions to the scalar field equation in the probe limit, we proceed to solving the full system, Eqs (\ref{FEqScalar}) -- (\ref{FEqa}).  As mentioned above, we solved the system numerically, and we now discuss the pattern of  solutions in the space of the parameters $\alpha$ and $Q$. We limit the presentation to solutions with no node of the function $\phi$ and insist on solutions with $\phi(r) \rightarrow 0$ rather than approaching a non-zero constant which exist as well.
Technically the solutions are constructed by fixing the parameter $\alpha$ and increasing gradually the value  $\phi(r_h)$; for a choice of these parameters
an hairy black hole (HBH) exists just for a very specific value of the charge, i.e. $Q = Q(\alpha, \phi(r_h))$.
The data corresponding to the typical value $\alpha=1$ is plotted on the left hand side of Fig. \ref{data_ga_3_bis}. We see that solutions exist from vanishingly small $\phi(r_h)$ up to a maximal value of $\phi(r_h)$ and for all solutions $Q>M$. The temperature decreases monotonically with $\phi(r_h)$, while the scalar charge $D$ vanishes at both ends and has a maximum at a certain point in the middle.

On the right hand side of Fig. \ref{data_ga_3_bis} three curves of the scalar charge $D$  versus the dimensionless ratio $Q/M$ appear for for $\alpha = 1, 3, 20$. The loop structure results from the non-monotonic dependence of $D$ and $Q/M$ on $\phi(r_h)$.

\begin{figure}[t]
\begin{center}
{\label{non_rot_cc_1}\includegraphics[width=7.4cm, angle = -00]{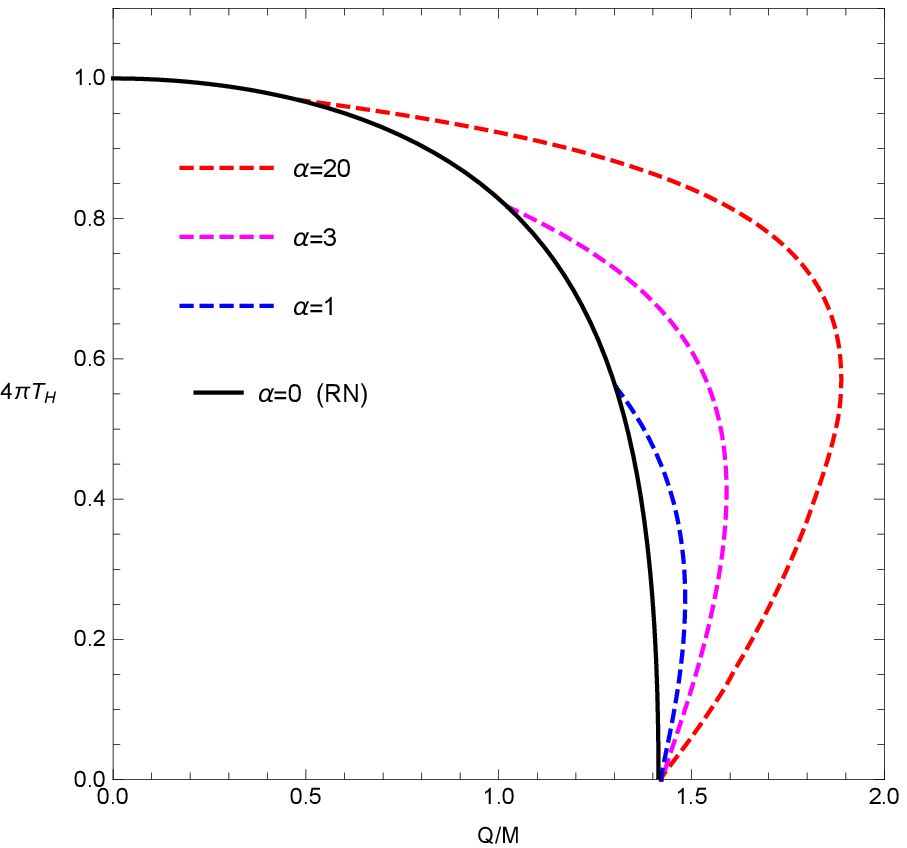}}
{\label{non_rot_cc_2}\includegraphics[width=8cm, angle = -00]{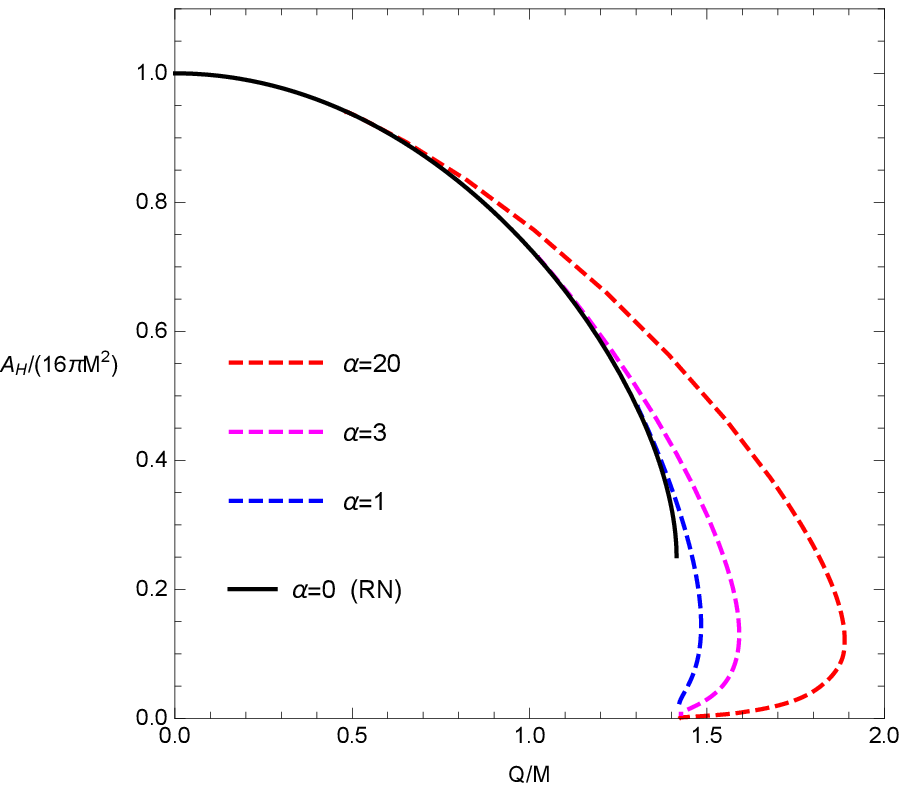}}
\end{center}
\caption{The temperature and horizon area as function of  $Q/M$  for the RN (solid lines) and hairy black holes with $\alpha=1, 3, 20$ (dashed lines). Left: Temperature.  Right: The ratio  $A_H/ 16 \pi M^2$.
\label{ThermodynData}
}
\end{figure}

The behavior of the temperature and horizon area of HBH is shown in Fig. \ref{ThermodynData}. The temperature and the reduced horizon area $a_h \equiv A_H/ 16 \pi M^2$ are plotted versus the dimensionless ratio $Q/M$ for $\alpha = 1, 3, 20$ as well as for the RN family ($\alpha=0$). The branches of HBH bifurcate from the RN curve respectively for $Q \approx 0.94$, $Q \approx 0.605$ and $Q \approx 0.26$.
The curves of $T_H$ and $a_h$ versus $Q/M$ actually represent two families of HBH,  merging at a maximal value of the parameter $Q/M$ which increases with $\alpha$. The solutions correspond to different intervals of $\phi(r_h)$: small and large. The solutions  with small $\phi(r_h)$ bifurcate from the RN
forming the upper branch (i.e. larger $T_H$ and $a_h$). The second branch corresponds to larger values of $\phi(r_h)$ (typically $\phi(r_h) \sim 3.0$)
and terminates at $Q/M = \sqrt{2}$, irrespectively of the value of coupling constant $\alpha$
(an explanation of this feature will come later).
The figure further reveals that, on the whole region where RN and HBH coexist,
the reduced horizon area $a_h \equiv A_H/ 16 \pi M^2$ is larger in the case of the hairy solutions.
Invoking the Bekenstein-Hawking black hole entropy formula,
it turns out that the HBH are {\it entropically preferred} with respect to RN black holes.
This feature seems to be common with  HBH constructed in other theories
(see e.g. refs. \cite{Silva:2017uqg,Herdeiro:2018wub,Fernandes:2019rez}).
This property is however not sufficient to guarantee the stability of HBH; this would need
a perturbative (or dynamical) analysis which is out of the scope of this paper.

\begin{figure}[t!]
\begin{center}
{\label{TwoProfiles}\includegraphics[width=8cm, angle = -00]{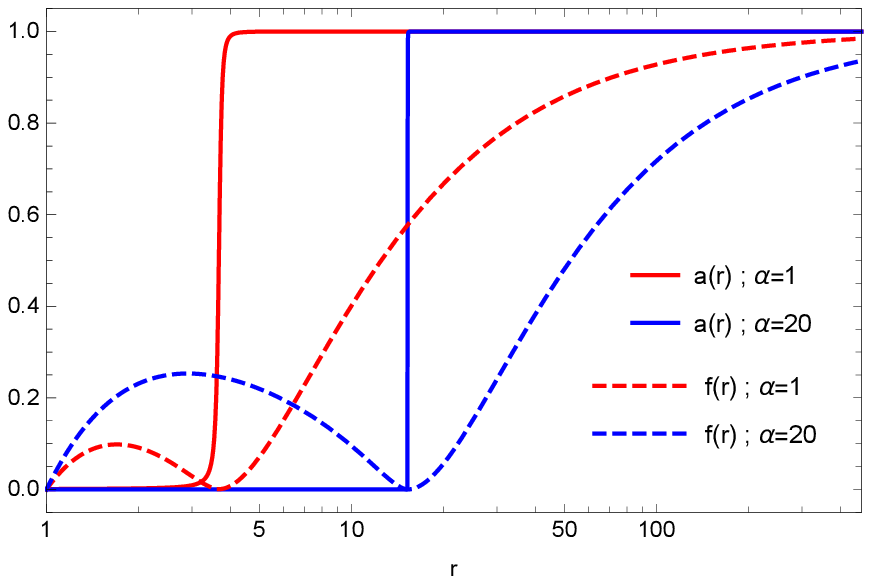}}
{\label{ParameterSpace}\includegraphics[width=8cm, angle = -00]{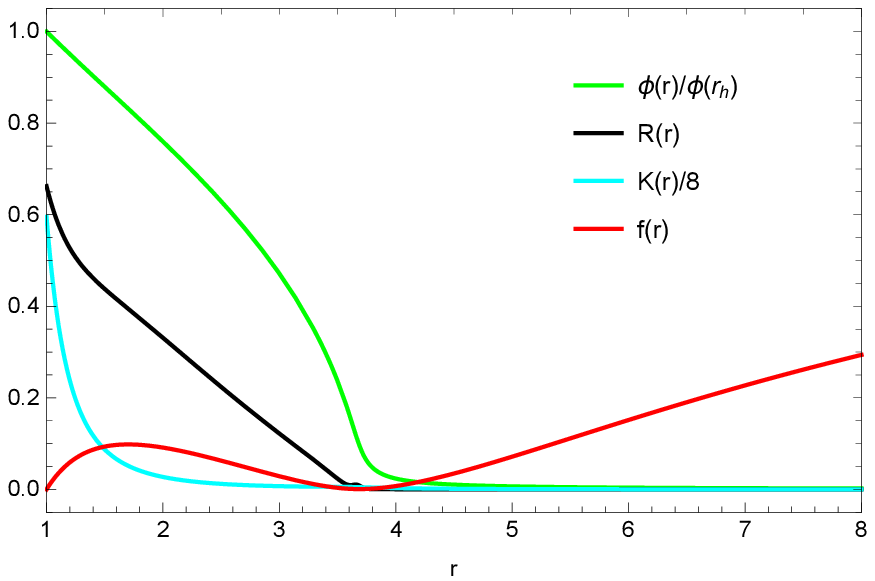}}
\end{center}
\caption{Left: Two profiles of the metric functions $f(r), a(r)$ for two critical HBH for $\alpha = 1, 20$ which turn into extremal RN solutions.
Right: Profiles of $f(r), \phi(r)$ and the Ricci and Kretschmann invariants for the limiting configuration
corresponding to $\alpha = 1$.
\label{profiles}
}
\end{figure}

Let us now discuss the  critical phenomenon limiting the families of HBH while increasing $\phi(r_h)$.
This can be understood from Fig.\ref{profiles}.
On the left, we show the metric functions $f(r), a(r)$ for the limiting configurations corresponding to
$\alpha = 1$ and $\alpha = 20$ for the value of $\phi(r_h)$ very close to the critical value.
The metric is characterized by two phenomena clearly appearing on the figure~:
\begin{itemize}
\item A second horizon as seen by a second zero of the  metric function $f(r)$ is approached  for some radius $r_c$ (depending on $\alpha$)
      such that $r_c > r_h$.
      The horizon $r_c$ is extremal. Outside this horizon, the scalar field tends to null function and
			the metric approaches an extremal RN black holes.
\item The function $a(r)$ is very small for $r \in [r_h, r_c]$ and raises steeply to $a(r)=1$ for $r > r_c$.
\end{itemize}
The fact that all branches of HBH solutions terminates at $Q/M \sim \sqrt 2$ on Fig.\ref{ThermodynData} is then
clear since the asymptotic  charges (Mass and electric charge)  of the HBH in this limit corresponds to the ones
of an extremal RN solutions. Furthermore it was checked numerically that the temperature of the HBH tends to zero in the critical limit.
Confirming the above statement, we present at the right hand side of Fig.\ref{profiles}
the profiles of the Ricci and Kretschmann invariants $R,K$ as well as the scalar field $\phi(r)$ for the case $\alpha=1$.
All these quantities are very close to zero for $r > r_c$. Furthermore, the invariants $R,K$ are finite everywhere
including the horizon $r_h$ in spite of the fact that the value $a(r_h)$ becomes very small. We noticed a similar phenomenon for charged boson stars in a scalar-tensor Horndeski theory of the ``John type'' \cite{Verbin-Brihaye2017} where in certain circumstances a boson star loses its scalar hair and turns into an extremal RN black hole.

It is worth emphasizing that the pattern of HBH in the present model is, up to our knowledge,  different from
any other  HBH present in the literature. In particular these are characterized by a single branch in the $Q/M, A_H/M^2$ plane 
and terminate into a singular configuration.

\section{Summary}\label{summary}
\setcounter{equation}{0}
We have discovered a new type of  scalarized charged black holes in a surprisingly simple system:
The Einstein-Maxwell-Klein-Gordon Lagrangian, supplemented by a non-minimal coupling
issued for the general vector-tensor theory formulated by G. Horndeski.
Several kinds of new compact objects, including soliton-like and black holes with scalar hair,
were constructed by using a spherically symmetric ansatz for the fields.
In particular, the families of hairy black holes are constructed  presenting different features
 from solutions constructed  in various existing extended gravity models.

This system deserves a further study in various directions like  the thermodynamic properties
of these new hairy black holes, their angular and/or radial excitations and  the
solutions with asymptotically non-vanishing scalar field which we found to exist.
 Other types of compact objects, like boson stars,
  exist also  when the scalar field becomes complex and massive.
These directions as well as others are currently under investigation.
\\
\\
\noindent {\bf Acknowledgments :} Y.B. gratefully acknowledges the Physics Division at the Open University of Israel in Raanana, for hospitality during an invited visit where  this work was initiated. Support from the Research Authority of the Open University of Israel is also acknowledged.



\vskip 1.5cm



 \end{document}